\documentclass{eas}
\usepackage{graphicx}
\begin{document}

\title{Multimode Radiative beta Cephei and delta Cephei Models} 
\author{Smolec R.}\address{Copernicus Astronomical Center, Bartycka 18, 00-716 Warsaw, Poland}
%
%
\begin{abstract}
As a by-product of $\beta$~Cephei and $\delta$~Cephei radiative model surveys we have found interesting cases of multimode pulsation. In case of $\beta$ Cephei stars, we found two multimode domains with two or three modes being involved. The origin of the multimode pulsation can be traced to one of the two different mechanisms: either to the non-resonant coupling of the two excited modes (double-mode pulsation) or to the $2\omega_1 \simeq \omega_0 + \omega_2$ parametric resonance (triple-mode pulsation). In case of $\delta$~Cephei models, a triple-mode pulsation domain, connected with the 2:1 resonance, $2\omega_0\simeq\omega_2$, is found. These models are of theoretical interest only, since they do not obey a proper mass-luminosity relation.
\end{abstract}
\maketitle
\section{Introduction}

Doubly-periodic classical pulsators (beat Cepheids, beat RR~Lyrae stars) are the simplest and one of the most important examples of multimode stellar pulsation. Observed stellar variation results from simultaneous oscillation in two low-order radial modes of pulsation. More than 30 years ago it was realized, that simple linear theory may be used to calculate the masses of such pulsators (Petersen \cite{JOP73}). However it took many years to understand and to model doubly-periodic pulsators. Some questions remain open even today.

Throughout this paper we will focus on doubly-periodic pulsations. We stress that doubly-periodic is not equivalent to double-mode, as more than two pulsation modes may be involved in doubly-periodic oscillations if resonant mode interaction is present. Such examples will be shown in this paper.

Attempts to model the doubly-periodic  pulsators started with early radiative lagrangian hydrocodes. These first nonlinear calculations failed to reproduce the doubly-periodic behaviour, but they were very incomplete, as direct nonlinear calculations are very time-consuming. Introduction and implementation of relaxation technique by Baker \& von Sengbusch (\cite {NBKS69}) and Stellingwerf (\cite{RS74}) allowed for efficient search for multimode solutions. Relaxation technique allows for fast convergence to limit cycle (periodic, finite-amplitude pulsation) and provides information on its stability through the Floquet stability coefficients (or so-called switching rates). Two Floquet coefficients are of interest, when one searches for doubly-periodic models with fundamental and first overtone modes simultaneously excited. $\gamma_{0,1}$ measures the stability of the first overtone limit cycle (switching rate toward fundamental mode), while $\gamma_{1,0}$ measures the stability of the fundamental mode (switching rate toward first overtone). If Floquet coefficients of both limit cycles are positive, neither of the limit cycles is stable and simultaneous pulsation in both modes is unavoidable. 

Using relaxation technique Kov\'acs \& Buchler (\cite{GKRB88}) performed detailed survey in search for the multimode pulsation in RR~Lyrae models. They found that the models close to the $2\omega_0\simeq\omega_3$ resonance center exhibit the doubly-periodic behaviour, however with periods inconsistent with observed ones. Physical reasons for resonant, doubly-periodic pulsations were analysed in detail by Dziembowski \& Kov\'acs (\cite{WDGK84}) through the analysis of the amplitude equations. Linearly damped third overtone acts as an energy sink for the linearly unstable fundamental mode. As a result amplitude of the fundamental mode is lowered and otherwise dominant fundamental mode is no longer able to saturate the pulsational instability. This allows the growth of the first overtone mode. Resulting triple-mode solution is apparently doubly-periodic due to nonlinear frequency-lock phenomenon. In the nonlinear, resonant regime $\tilde{\omega}_3$ is exactly equal to $2\tilde{\omega}_0$ and third overtone is not visible as a separate frequency, but contributes to the harmonic frequencies of the fundamental mode.  Other examples of resonant triple-mode, doubly-periodic solutions will be presented in the next Sections.

Kov\'acs \& Buchler (\cite{GKRB88}) also found non-resonant double-mode models through playing with artificial viscosity parameters. Artificial viscosity provides unphysical dissipation  necessary to handle the shocks in the radiative codes. Model properties, specially the amplitudes depend on artificial viscosity (eg. Kov\'acs \cite{GK90}). With artificial viscosity parameters chosen to satisfy the observed amplitudes no multimode solution was found. However models that are less dissipative due to reduced artificial viscosity, exhibit steady doubly-periodic, double-mode pulsation as was investigated in detail by Kov\'acs \& Buchler (\cite{GKRB93}). Their models however, strongly depend on numerical details and do not reproduce all the observational constraints. Nevertheless, experiments with artificial viscosity motivated the development of numerically less dissipative codes with physical dissipation instead of artificial one. The success came with the introduction of turbulent convection energy transfer recipies into the hydrocodes. Koll\'ath et al. (\cite{ZKea98}) reported double-mode Cepheid models, while double-mode RR~Lyrae models were reported by Feuchtinger (\cite{MF98}). These double-mode models originate from non-resonant coupling of the pulsation modes and reproduce the observed periods and period ratios.

In this paper we present new interesting examples of purely radiative doubly-periodic models. In Section~2 we discuss numerically robust, resonant and non-resonant $\beta$~Cephei models, while in Section~3 resonant $\delta$~Cephei models will be presented.

\section{Multimode radiative $\beta$ Cephei models}

$\beta$~Cephei stars are main sequence, early B-type pulsators. Acoustic modes and low-order gravity modes are driven through the $\kappa$-mechanism acting in the iron opacity bump (e.g. Pamyatnykh \cite{AP99}). Most of these stars are multiperiodic. Non-radial modes dominate, although radial pulsation is observed in several stars. Under some simplified assumptions Smolec \& Moskalik (\cite{RSPM07}) studied the amplitude saturation in these stars, using purely radial, radiative codes. In case of these stars convection may be safely disregarded, as $\beta$ Cephei stars are too hot to have an effective energy transfer by convection (Pamyatnykh \cite{AP99}, Smolec \& Moskalik \cite{RSPM07}). Model properties are also independent of artificial viscosity, contrary to the classical pulsators case (see e.g. Kov\'acs \cite{GK90} and Smolec \& Moskalik \cite{RSPM07}). Thus, calculated radiative models are very robust. As a by-product of this analysis several multimode models were found. The modal selection depends on the opacities being used. Here we discuss the models computed with the OPAL opacities, as they represent more interesting and clear case. Multimode models calculated with the OP opacities are discussed in detail by Smolec \& Moskalik (\cite{RSPM07}).

\begin{figure}[h]
\includegraphics[width=8cm]{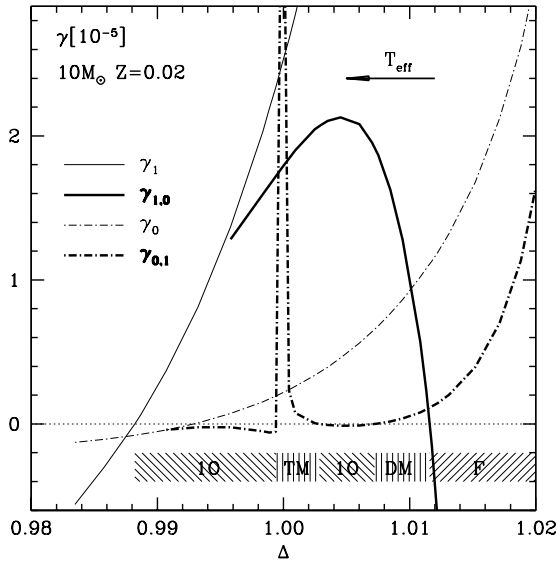}
\caption{Linear growth rates (thin lines) and Floquet coefficients (thick lines) plotted versus the $\Delta$ parameter ($\Delta=2\omega_1/(\omega_0+\omega_2)$) for the sequence of $\beta$~Cephei models. Shaded bar indicates the modal selection as derived from Floquet coefficients (1O - first overtone, F - fundamental mode, TM - triple-mode, DM - double-mode pulsation domains).}
\label{fig1}
\end{figure}

Computation of the Floquet stability coefficients for the models located on 10M$_\odot$ evolutionary track revealed two multimode domains where first overtone and fundamental mode are simultaneously unstable (both Floquet coefficients, $\gamma_{0,1}$ and $\gamma_{1,0}$, are positive). Linear frequencies of the models show, that one of the domains is clearly connected with the high-order parametric resonance $2\omega_1\simeq\omega_0+\omega_2$. Situation is illustrated in Fig.~\ref{fig1}, where Floquet coefficients and linear growth rates of the modes, $\gamma_0$ and $\gamma_1$, are plotted versus the $\Delta$ parameter, that characterizes the proximity of the resonance, $\Delta=2\omega_1/(\omega_0+\omega_2)$. Starting from the blue edge, switching rate toward fundamental mode is negative and thus, first overtone is stable. Then, exactly at the resonance center, a prominent peak of the switching rate toward fundamental mode is visible. Otherwise negative, $\gamma_{0,1}$ becomes positive at the resonance center. Thus, the resonance leads to the destabilization of the first overtone limit cycle. Resulting triple-mode domain appears as an island in the middle of the first overtone pulsation domain. The resonance does not affect the stability of the fundamental mode, as is well visible in Fig.~\ref{fig1}, and as was shown by Kov\'acs \& Buchler (\cite{GKRB93}) through the analysis of the amplitude equations. Second multimode domain located at $\Delta\approx 1.01$ (Fig.~\ref{fig1}) is not connected with any resonance. Thus, only fundamental and first overtone modes are involved in double-mode oscillations. This domain separates the first overtone and fundamental mode pulsation domains as is expected in case of non-resonant mode coupling. 

\begin{figure}[h]
\includegraphics[width=8cm]{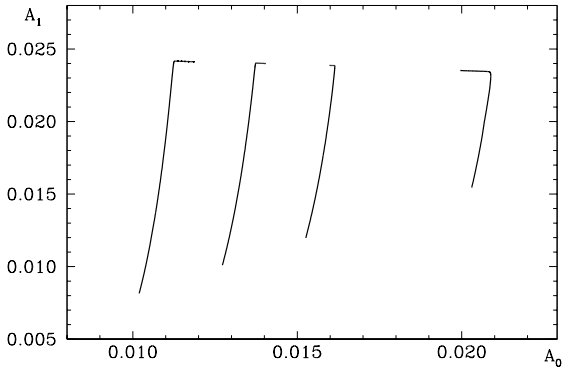}
\caption{Temporal evolution of the first overtone, $A_1$, and the fundamental mode, $A_0$, amplitudes of the non-resonant $\beta$~Cephei model. Integrations were initialized with four different initial conditions and followed for more than hundred thousand cycles.}
\label{fig2}
\end{figure}

We have also performed some nonlinear calculations to approach the doubly-periodic solution. Due to the very low growth rates, several hundred thousand cycles are necessary in nonlinear integrations to approach the attractor. Thus, we focused on only one model located in the non-resonant double-mode domain. Nonlinear integrations were initiated with four different initial conditions (velocity perturbations). Relative surface displacement, $\delta R/R$, was then followed and the amplitudes of the fundamental and first overtone modes were extracted through the use of the analytic signal method (Koll\'ath \& Buchler \cite{ZKRB01}). Fig.~\ref{fig2} presents the results in the modal amplitude phase-space. Transient evolution is extremely slow and none of the trajectories reached the double-mode attractor. However its location is clearly visible: two trajectories evolve toward the attractor from the right, and two of them from the left.  

Both of the discussed multimode domains represent clear examples of radial multimode pulsation. Origin of the multimode state has clear physical interpretation. Computed models are robust, independent of numerical details. Also radiative codes are fully applicable in case of $\beta$~Cephei stars. Nevertheless, two radial modes simultaneously excited, were not observed in these stars up to date. These results however, shed some light on the reasons of multiperiodic nature of $\beta$~Cephei stars. Saturation properties of the non-radial acoustic modes should be very similar to the properties of the radial modes. Thus, non-resonant coupling of non-radial modes may lead to multimode pulsations. Taking into account the dense spectrum of the non-radial modes, resonant coupling leading to multimode pulsation is also very likely (Dziembowski \cite {WD07}, Smolec \& Moskalik \cite{RSPM07}). 

\section{Triple-mode radiative $\delta$ Cephei models}

Triple-mode models we will discuss in this Section, are connected with the 2:1 resonance between the linearly unstable fundamental mode and the linearly damped second overtone mode, $2\omega_0\simeq\omega_2$. This resonance plays a crucial role in shaping the Hertzsprung bump progression in fundamental mode Cepheids (Kov\'acs \& Buchler \cite{GKRB89}; Buchler, Moskalik \& Kov\'acs \cite{RBea90}). These so-called bump Cepheids are indeed a double-mode pulsators. However, due to frequency-lock phenomenon, second overtone is not visible as a separate frequency, but manifests as a nonlinear distortion of the light/velocity curves. Observationally, resonance center falls at fundamental mode periods around $P_0\approx 10$ days. Due to the resonant interaction with the damped second overtone, amplitude of the fundamental mode is lowered, which may lead to the destabilization of its limit cycle. Indeed, Buchler et al. (\cite{RBea97}) show that for some models, $\gamma_{1,0}>0$, close to the resonance center and thus, fundamental mode becomes unstable. However, these models are located relatively close to the blue edge and the first overtone is stable ($\gamma_{0,1}<0$). Thus, the star pulsates in the first overtone, and simultaneous pulsation in fundamental and first overtone mode is not possible. This is in agreement with observations, since double-mode first overtone/fundamental mode Cepheids have much shorter periods and, as linear calculations show, lie in a non-resonant regime.

\begin{figure}[htpb]
\includegraphics[width=12.5cm]{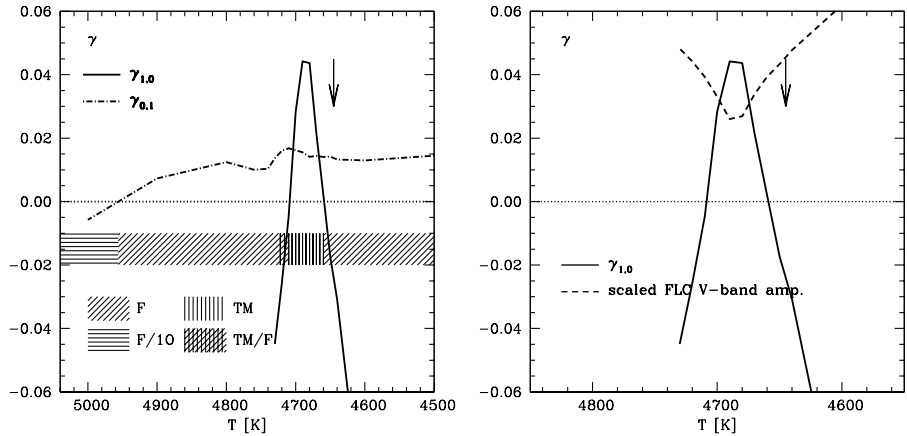}
\caption{Floquet coefficients plotted versus the model's temperature (left panel). Shaded bar indicates modal selection derived from Floquet coefficients and from nonlinear integrations. Destabilization of the fundamental mode limit cycle ($\gamma_{1,0}>0$) is clearly connected with its lower amplitude (right panel). Arrows indicate the position of the $2\omega_0\simeq\omega_2$ resonace center.}
\label{fig3}
\end{figure}

Using different mass-luminosity relation we have shifted the resonance center toward much lower temperatures and longer fundamental mode periods. As a result resonance center falls at the temperature range, where first overtone is already unstable, $\gamma_{0,1}>0$. Situation is illustrated in the left panel of Fig.~\ref{fig3}. For higher temperatures both limit cycles are stable. Thus, pulsation in either fundamental mode or first overtone is possible. For lower temperatures first overtone limit cycle becomes unstable and fundamental mode pulsation domain begins. Then, close to the resonance center, otherwise stable, fundamental mode becomes unstable. This is a direct consequence of lower amplitude of fundamental mode, as is shown in right panel of Fig.~\ref{fig3}. Amplitude is lowered because part of the fundamental mode energy is resonantly transferred to the damped second overtone. Triple-mode pulsation, with two frequencies (second overtone is phase-locked) is possible. We stress that these models are of theoretical interest only, as they result from arbitrary mass-luminosity relation, that is inconsistent with observations.

\begin{figure}[t]
\includegraphics[width=7.5cm]{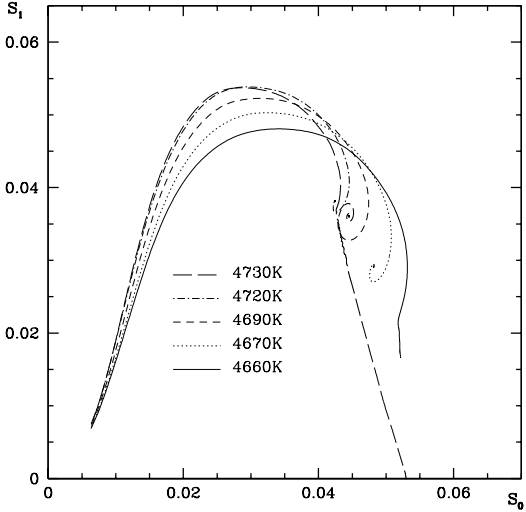}
\caption{Phase-space plots for models with different temperatures. All integrations were started with the same initial condition and continued till the attractor was reached.}
\label{fig4}
\end{figure}

\begin{figure}[t]
\includegraphics[width=12.5cm]{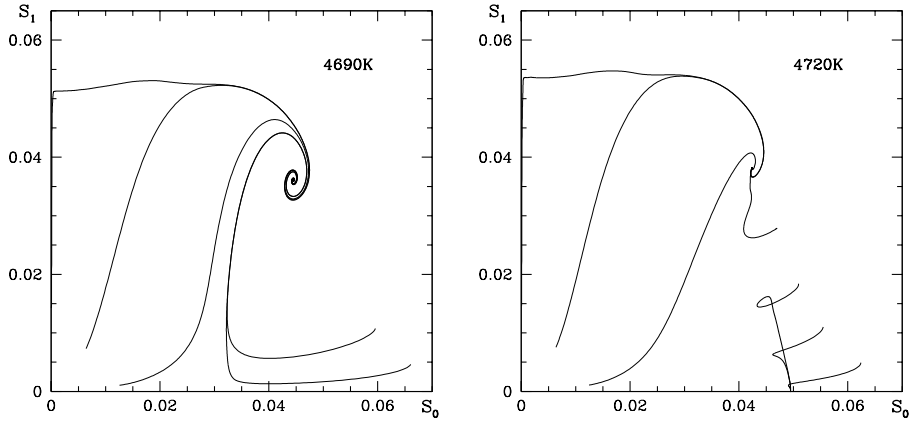}
\caption{Structure of the phase-space for 4690K model (left panel) and 4720K model (right panel). In the latter case two attractors are visible.}
\label{fig5}
\end{figure}

Several nonlinear integrations were performed to approach the triple-mode attractor. Relative surface displacement of the model, $\delta R/R$, was followed. Analytic signal method was used to extract the amplitudes, $S_0$ and $S1$, corresponding to the variability with frequencies $\omega_0$ and $\omega_1$. $S_1$ is directly related to the amplitude of the first overtone pulsation, $S_1=A_1$. However, due to resonant coupling between the fundamental and second overtone modes, $S_0$ contains contribution from both of the modes, namely:
$$S_0=\Re (a_0+h_{21}a_0^*a_2)$$
where $a_0$ and $a_2$ are complex amplitudes corresponding to fundamental and second overtone modes respectively and $h_{21}$ is complex coefficient (Kov\'acs \& Buchler \cite{GKRB89}). We note that the topology of the $S_0$-$S_1$ phase-space is the same as the topology of the $A_0$-$A_1$ phase-space. Exemplary trajectories in $S_0$-$S_1$ phase-space are shown if Figs.~\ref{fig4} and \ref{fig5}. In Fig.~\ref{fig4} we show the trajectories for five models with different temperatures. Each model on this Figure was kicked with the same initial condition. We also investigated the details of the phase-space by integrating the same static model with different initial conditions. Results are shown in Fig.~\ref{fig5} for two different models: 4690K model (left panel) and 4720K (right panel). Several interesting features are visible in Figs.~\ref{fig4} and \ref{fig5}. 

(i) The 4730K model trajectory evolves toward single-frequency solution. At lower temperature (4720K) triple-mode attractor appears. Trajectories for the 4690K model (Figs.~\ref{fig4} and \ref{fig5} left) evolve toward a pronounced spiral attractor. Such attractor is not possible in non-resonant case. For lower temperatures spiral becomes less pronounced and disappears for 4660K model. For slightly cooler model triple-mode attractor also disappears. 

(ii) For lower temperatures triple-mode solution is the only attractor of the system (Fig.~\ref{fig5} left). For these models both Floquet coefficients are positive (Fig.~\ref{fig3} left). 

(iii) For higher temperatures (Fig.~\ref{fig5} right) hysteresis is possible. Depending on the initial conditions, trajectories evolve toward two different attractors: a triple-mode attractor and a single-frequency fundamental mode attractor. For these models fundamental mode is stable ($\gamma_{1,0}<0$, Fig.~\ref{fig3}). Similar hysteresis (double-mode/fundamental mode) is present in non-resonant convective Cepheid and RR~Lyrae models (Koll\'ath \& Buchler \cite{ZKRB01}). However here, the single-frequency attractor is a double-mode one, being a bump Cepheid.

{\bf Acknowledgements}. I am grateful to Pawe\l{} Moskalik for reading and commenting the manuscript. I wish to thank the organizers for the possibility to attend the meeting and to give a talk. This work has been supported by the Polish MNiI Grant No. 1 P03D 011 30. 



\end{document}